\documentclass[11pt,a4paper]{article}
\usepackage{jheppub}
\newif\ifonecol\onecoltrue
\pdfoutput=1

\usepackage{amsmath}
\usepackage{amssymb}
\usepackage{amsfonts}
\newcommand{\doctitle}{Reconciling neutrino flux from heavy dark matter decay and recent events at IceCube}

\usepackage{hyperref} 
\usepackage[all]{hypcap}
\usepackage{graphicx}
\usepackage{array}
\usepackage{slashed} % For fermionic slash notation
\usepackage{hyphenat}
\usepackage{dcolumn}
\usepackage{subcaption}
\usepackage{dcolumn} % Align table columns on decimal point
\usepackage{paralist}
\usepackage{mathrsfs} % Script fonts in math mode

\newcommand{\ie}{i.e.}
\newcommand{\eg}{e.g.}

\newcommand{\anti}[1]{\ensuremath{\bar{#1}}}

\newcommand{\dm}{\ensuremath{\text{DM}}}
\newcommand{\vhdm}{VHDM}
\newcommand{\sm}{SM}

\newcommand{\order}[1]{\ensuremath{\mathcal{O}(#1)}}
\newcommand{\viz}{\textit{viz.}}
\newcommand{\nue}{\ensuremath{\nu_e}}
\newcommand{\numu}{\ensuremath{\nu_\mu}}
\newcommand{\nutau}{\ensuremath{\nu_\tau}}

\newcommand{\astro}{\ensuremath{\text{Astro}}}
\newcommand{\ic}{\ensuremath{\text{IC}}}
\newcommand{\cmssr}{\ensuremath{\text{ cm}^{-2}\text{ s}^{-1}\text{ sr}^{-1}}}
\newcommand{\flxunit}{\ensuremath{\text{ GeV}\cmssr}}
\newcommand{\chisq}{\ensuremath{\chi^{2}}}
\newcommand{\fgal}{\ensuremath{\Phi^\text{G}}}
\newcommand{\fxgal}{\ensuremath{\Phi^\text{EG}}}
\newcommand{\ud}{\ensuremath{\mathrm{d}}}
\newcommand{\fdnde}{\ensuremath{\frac{\ud N_{\nu}}{\ud E_{\nu}}}}
\newcommand{\dnde}{\ensuremath{\ud N_{\nu}/\ud E_{\nu}}}
\newcommand{\diff}{\ensuremath{\,\ud}}
\newcommand{\rsun}{\ensuremath{R_{\odot}}}

\newcommand{\cl}{C.L.}

\usepackage{color}
\definecolor{deepred}{RGB}{210,0,0}
\newcommand{\edited}[1]{#1}
\newcommand{\newinsrt}[1]{#1}

\title{\doctitle}

\author[a]{Atri Bhattacharya,}
\author[b]{Mary Hall Reno}
\author[a,c]{and Ina Sarcevic}
\affiliation[a]{Department of Physics, University of Arizona, Tucson, AZ 85721, USA}
\affiliation[b]{Department of Physics and Astronomy, University of Iowa, Iowa City, IA 52242, USA}
\affiliation[c]{Department of Astronomy and Steward Observatory, University of Arizona, Tucson, AZ 85721, USA}
\emailAdd{atrib@email.arizona.edu}
\emailAdd{mary-hall-reno@uiowa.edu}
\emailAdd{ina@physics.arizona.edu}

\abstract{The IceCube detector has recently reported the observation of 28 events at previously unexplored energies.
While the statistics of the observed events are still low, these events hint at the existence of a neutrino flux over and above the atmospheric neutrino background.
We investigate the possibility that a significant component of the additional neutrino flux originates due to the decay of a very heavy dark matter (VHDM) particle via several possible channels into standard model particles.
We show that a combination of a power law astrophysical neutrino spectrum and the neutrino flux from the decay of a \dm\ species of mass in the range $150-400$ TeV improves the fit to the observed neutrino events than that obtained from a best-fit astrophysical flux alone.
\edited{Assuming the existence of an astrophysical background described by the IC best-fit, we also show that, for the decay of even heavier \dm\ particles ($m_{\text{DM}} \sim 1$ PeV), the same observations impose significant constraints on the decay lifetimes.}
\newinsrt{Allowing the astrophysical flux normalization to vary leads to modifications of these limits, however, there is still a range of dark matter mass and lifetime that is excluded by the IC results.}
}


\arxivnumber{1403.1862}

\begin{document}
\maketitle

\section{Introduction}
The IceCube detector (IC) at the South Pole has recently observed 28 events at energies 30 TeV -- 1.2 PeV \citep{Aartsen:2013jdh}.
This marks the first ever observation of neutrino events at PeV energies in any detector and heralds a new chapter in understanding the physics responsible for producing and accelerating fundamental particles to such high energies.
The commonly accepted means for the production of fundamental particles at these so-called ultra-high energies (UHE) is at the extremely reactive cores and jets of astrophysical sources \cite{Waxman:1995dg} beyond our galaxy, where the already fast-moving charged particles can be accelerated further due to Fermi acceleration \cite{Fermi:1949ee} when traversing through the shock waves associated with these sources.
Standard theoretical arguments \cite{PhysRevD.59.023002} deduce that the diffuse flux of neutrinos arriving at the Earth from the such astrophysical sources should have a power-law energy spectrum proportional to $E^{-2}$.
\newinsrt{While several possible astrophysical sources have been suggested \cite{Essey:2009ju, Essey:2010er, Kalashev:2013vba, Murase:2013rfa, Murase:2013ffa} to explain these observations, given the present low nature of the statistics, it is difficult to determine the origin of these high energy neutrinos with strong statistical significance.}
\edited{The recent observations at IC hint at possibly first detection of a diffuse neutrino astrophysical flux;} the power-law spectrum ($\diff\Phi/\diff E \propto E^{-\alpha}$) is consistent with $\alpha = 2.0$.
However, there are some noteworthy incompatibilities between the observed events and the corresponding $E^{-2}$ power-law spectrum: 
\begin{inparaenum}[\itshape a\upshape)]
	\item the expected number of events in the lower energies 30--150 TeV falls consistently below the observations, and
	\item at two high energy bins marked by a distinct lack of observed events, the $E^{-2}$ flux with the best-fit normalization obtained by IC gives event rates that are simply too high.
\end{inparaenum}

While it might be entirely possible, indeed even likely, given the small sampling of data we are working with, that these differences are merely statistical fluctuations which will disappear with the accumulation of more data, the systematic nature of the difference between theory and experiment, especially at the lower energy bins, suggests that the total neutrino flux incident at the Earth might comprise a component additional to that from astrophysical sources.
Given the extremely high energies in question, the sources that can produce a neutrino flux that should be detectable at IC are rather few.
One interesting source, apart from astrophysical ones, that is capable of producing UHE neutrinos is  from the interactions or decays of a very heavy dark matter (\vhdm) species.

The fact that a sizable fraction, about 27\%, of the universe's energy density is made of non-luminous matter is now well known, and has been recently confirmed by the PLANCK observations \cite{arXiv:1303.5076}.
The most consistent explanation of this large amount of energy not accounted for by standard baryonic matter is given by postulating the existence of one or more, usually heavy, dark matter particles which are relatively inert to interactions with the standard model particles (see, \eg, \cite{Bertone:2004pz}).
Efforts to experimentally detect such particles, called Weakly Interacting Massive Particles (WIMP), are ongoing \cite{Ahmed:2009zw, Bravin:1999fc, Aprile:2011hi, Armengaud:2011cy, Bernabei:2010mq, Archambault:2012pm}.
While these direct search experiments are constrained to focus on detecting WIMP-like particles with masses in the range 10 GeV -- 1 TeV, \dm\ species in nature might be significantly heavier.
For thermal relics, the mass is constrained by unitarity to lie \edited{in the low TeV mass range} \cite{Griest:1989wd, Chung:1998zb}; however, nothing prevents non-thermal \dm\ masses to be even as high as a $10^{9}$ TeV \cite{Chung:1998zb}.
Indirect searches such as the detection of products of \dm\ annihilation in the sun at large volume detectors like the IC are sensitive to heavier \dm\ species with masses above the TeV's as well.

In this paper, without going into the specific models that predict the existence of such a candidate, we assume the existence of a \dm\ species in the universe with its mass in the range 100 TeV -- 1 PeV.
If it exists, both annihilation and/or decay (if it is unstable) of such a \vhdm\ particle, will lead to the production of standard model particles eventually generating neutrinos as secondaries.
In this paper, we will investigate the possibility that neutrinos from the decay of such a species might be able to assuage some of the tension between observation and theoretical predictions as outlined in the previous paragraph.

Specifically we will focus on a \vhdm\ species which is unstable, and can, therefore, decay but with lifetimes well beyond the age of the universe --- present constraints from gamma ray observations tie the \dm\ lifetimes to $\gtrsim 10^{26}$ s \cite{Cirelli:2009dv, Murase:2012xs}.
In general the \dm\ could decay into different possible \sm\ final states, including charged leptons, neutrinos, quarks and weak interaction bosons W and Z.
Limiting ourselves to two-body decays, we will do a generic model independent study by analyzing the spectrum of neutrinos produced in each of the possible decay channels individually.
For each of the channels, we will evaluate the number of events expected at the IC from a total flux comprising neutrinos from \dm\ decay and an astrophysical power-law flux.
Finally we will show that, for a reasonable set of parameters for each decay mode, the event rates from the total (\dm\ + astrophysical) diffuse flux improves the statistical match to the observed events over that from a power-law spectra of the astrophysical flux alone.
In doing so, we will determine the \dm\ mass and lifetimes that lead to event rates most closely in match with the IC observations.

The paper is organized as follows. 
In section \ref{sec:dmdecay} we discuss models of decaying dark matter in existing literature and we calculate the resulting neutrino flux for the various decay channels.
For each decay mode in the different models, we calculate the resulting neutrino flux expected at Earth after standard oscillation amongst the three flavors.
In section \ref{sec:statanalysis}, we compute the net all-sky neutrino flux incident at IC by combining the flux from \dm\ decay obtained in the previous section with an astrophysical power-law spectrum and use this flux to calculate the event rates expected at IC.
In section \ref{subsec:ltbounds} we do a full parameter-space scan in the \dm\ mass-lifetime plane to show the region of the plane that is excluded by recent observations.
By doing \edited{suitable statistical analyses} of the predicted event rates against observations at IC, in section \ref{subsec:bfit} we derive the values of the \dm\ mass and lifetime that give the closest match.
Finally, we draw our conclusions in section \ref{sec:conc}.

\section{\label{sec:dmdecay}Dark matter decay modes and the resulting neutrino flux}
\dm\ candidates in several models are unstable and can decay to standard model particles. 
Decaying dark matter is a favored solution for the explanation of the $e^{\pm}$ excess in cosmic radiation seen by PAMELA \cite{Adriani:2008zr} and ATIC \cite{Chang:2008aa}.
Indeed, a certain class of \dm\ models, called leptophilic \dm\, in which the \dm\ particles annihilate or decay preferably to the charged leptons rather than to quarks, was first proposed \cite{Fox:2008kb} to explain these excesses.
Specifically, in the case of decay \cite{Yin:2008bs,Ibarra:2008jk,Nardi:2008ix}, the \dm\ particle can produce an excess of leptons via $\dm \to \ell^{+}\ell^{-}$, where $\ell$ is one of $e$, $\mu$ or $\tau$.
Decays to lighter charged leptons are generally more strongly constrained from the positron excess measurements by Fermi Large Area Telescope \cite{FermiLAT:2011ab}, and AMS-02 \cite{Aguilar:2013qda}, however, and decays to $e^{-}e^{+}$ pairs are strongly disfavored (see, \eg, \cite{Cirelli:2009dv}).
\dm\ particles might also decay to neutrinos producing a sharp resonance in the spectrum at $E_{\nu} = m_\dm / 2$ \cite{Feldstein:2013kka, Esmaili:2013gha}.
In addition heavy \dm\ candidates from supersymmetry such as the gravitino ($\psi_{3/2}$) can also decay, \eg, to $\gamma\nu$, but more preferably to $W^{\pm}\ell^{\mp}$ and $Z^{0}\nu$.

Here, we do a largely model-independent study of dark matter decay in the context of the recent events seen at IC, restricting ourselves to studying two-body decays of a bosonic \dm.
We consider each of the decay channels
\begin{inparaenum}[\itshape a\upshape)]
	\item $\dm \to \tau^{+}\tau^{-}$,
	\item $\dm \to \mu^{+}\mu^{-}$,
	\item \label{item:dmtoww}$\dm \to W^{+}W^{-}$, and
	\item \label{item:dmtozz}$\dm \to Z^{0}Z^{0}$,
\end{inparaenum}
and, in each case, compute the energy spectrum, \dnde, of neutrinos produced due to the decay of a \dm\ particle by using the event generator PYTHIA 8.1 \cite{Sjostrand:2007gs, Sjostrand:2006za}, taking care to include electroweak corrections \cite{Ciafaloni:2010ti}.
We do not consider decays to $e^{\pm}$ pairs because these are much more strongly constrained than any of the four channels considered above.
We also do not consider decays to $q\anti{q}$ pairs as the secondary neutrino flux produced when the \dm\ particle decays to quarks in the relevant energy range is low, and does not give observable event rates at the IC.
Neutrinos are produced as secondaries in all the four channels, in the case of \textit{\ref{item:dmtoww}}) and \textit{\ref{item:dmtozz}}) due to the fragmentation of the Z and W bosons.

The total neutrino flux from the decay of \dm\ is composed of contributions from interactions within the galactic halo and from outside the galaxy \cite{Esmaili:2012us}, \ie,
\begin{equation}\label{eqn:totdifff}
\frac{\ud\Phi}{dE} = \frac{\ud\fgal}{dE} + \frac{\ud\fxgal}{dE},
\end{equation}
where, $\fgal$ and $\fxgal$ represent the galactic and extra-galactic components of the total neutrino flux respectively.
The differential flux from \dm\ decay in the Milky Way halo is given by
\begin{equation}
\frac{\ud\fgal}{\ud E_{\nu}} = \frac{1}{4\pi\,m_\dm\,\tau_\dm} \fdnde \int^{\infty}_{0} \rho(r(s, l, b))\diff s,
\end{equation}
where, $m_{\dm}$ and $\tau_{\dm}$ represent the mass and lifetime of the dark matter particle, $l$ and $b$ represent the galactic coordinates of the place where the interaction occurs and $\rho$ represents the density profile of dark matter within the galaxy. The integral is over the line-of-sight parameter $s$, and it it related to the distance from the galactic centre $r$ by
\[r(s,l,b) = \sqrt{s^{2} + \rsun^{2} - 2 s \rsun \cos(b)\cos(l)},\]
\rsun\ being the distance of the sun from the GC, $\rsun = 8.5$ kpc.
Using the NFW \cite{Navarro:1996gj} profile for the dark matter distribution in the galaxy, the all-sky differential neutrino flux from \dm\ decay in the galaxy is given by (see, \eg, \cite{Esmaili:2012us}, \cite{Bai:2013nga})
\begin{equation}\label{eqn:difffgal}
\frac{\ud\fgal}{\ud E} = D_\text{G} \fdnde,
\end{equation}
where, \[D_\text{G} = 1.7\times 10^{-8} \left( \frac{1\text{ TeV}}{m_\dm} \right)
\left( \frac{10^{26}\text{ s}}{\tau_{\dm}} \right) \cmssr.\]
Additionally, the extra-galactic component of the differential neutrino flux is given by the expression
\begin{subequations}
\ifonecol
\begin{align}
\frac{\ud \fxgal}{\ud E} &= \frac{\Omega_{\dm}\,\rho_{\text{c}}}{4\pi\,m_\dm\,\tau_\dm}
\int^{\infty}_{0} \frac{1}{H(z)} \fdnde \left[(1+z)E_{\nu}\right]\diff z\\
\label{eqn:difffxgal}
&= D_\text{EG} \int^{\infty}_{0} \frac{1}{\sqrt{\Omega_{\Lambda} + \Omega_\text{m}(1+z)^{3}}} \fdnde \left[(1+z)E_{\nu}\right]\diff z,
\end{align}
\else
\begin{align}
\frac{\ud \fxgal}{\ud E} &= \frac{\Omega_{\dm}\,\rho_{\text{c}}}{4\pi\,m_\dm\,\tau_\dm}
\int^{\infty}_{0} \frac{1}{H(z)} \fdnde \left[(1+z)E_{\nu}\right]\diff z\\
\label{eqn:difffxgal}
&= D_\text{EG} \int^{\infty}_{0} \frac{1}{\sqrt{\Omega_{\Lambda} + \Omega_\text{m}(1+z)^{3}}}\nonumber\\
&\quad\quad\quad\quad\quad \times\fdnde \left[(1+z)E_{\nu}\right]\diff z,
\end{align}
\fi
\end{subequations}
with\[D_\text{EG} = 1.4 \times 10^{-8} \left( \frac{1\text{ TeV}}{m_\dm} \right)
\left( \frac{10^{26}\text{ s}}{\tau_{\dm}} \right) \cmssr.\]
Here, $z$ represents the red-shift of the source, $\rho_\text{c} = 5.6\times 10^{-6}\text{ GeV cm}^{-3}$ denotes the critical density of the universe, and we have used $H(z) = \sqrt{\Omega_{\Lambda} + \Omega_\text{m}(1+z)^{3}}$, and $\Omega_\Lambda = 0.6825$, $\Omega_\text{m} = 0.3175$, $\Omega_\dm = 0.2685$ and $H_{0} = 67.1 \text{ km}\text{ s}^{-1}\text{ Mpc}^{-1}$ from the recent PLANCK data \cite{arXiv:1303.5076}.

Having already computed the neutrino spectrum per decay, \dnde, we can now directly use it in eqns.\ \eqref{eqn:totdifff}, \eqref{eqn:difffgal} and \eqref{eqn:difffxgal} to calculate the total neutrino flux produced due to each decay mode of the \dm.

\subsection*{Effect of neutrino oscillation}
As the neutrinos propagate to the Earth, the three flavors of neutrinos oscillate amongst themselves, leading to an averaging out of the flux ratios \cite{Learned:1994wg,Athar:2000yw}. Decays to any of the gauge boson leads to the production of an equal number of neutrinos of each flavor at source.
Due to the effect of oscillation, a flux ratio of $\nue:\numu:\nutau = 1:1:1$ remains unchanged while propagating to the Earth.
\edited{On the other hand the decay of \dm\ to $\mu^\pm$ pairs leads to a neutrino flux of $1:1:0$ at source, which gets modified to $0.785:0.607:0.607$ due to neutrino oscillation during propagation, while decays to $\tau^{\pm}$ pairs lead to different flavor ratios based on the decay channel of the $\tau$.
Since, as is evident, flavor ratios at source may differ considerably depending on the \dm\ decay channel, we calculate the final flavor ratio by properly incorporating in our code the effect of oscillation as the neutrinos produced in a specific favor ratio propagate to Earth.
This is done for neutrino fluxes from the \dm\ decay as well as that from astrophysical sources. For the latter we assume the neutrinos are produced in a 1:2:0 flavor ratio consistent with that from pion decay.}

\section{\label{sec:statanalysis}Event rates at IC from DM decay and astrophysical neutrino fluxes}
\edited{The astrophysical neutrino flux arrives at the Earth essentially in a $1:1:1$ flavor ratio}\footnote{\edited{This is typically the case when they are produced in a $1:2:0$ ratio at the source, due to, \eg, $p\gamma$ or $pp$ interactions. We note that if one uses the best fit oscillation parameters \cite{GonzalezGarcia:2012sz} this ratio becomes $1.063:1:0.991$.
Using this exact ratio has only a minimal effect on our results.}} and follows an unbroken power law, \ie,
\begin{equation}
\frac{\diff\Phi_\astro}{\diff E} \equiv \frac{\diff}{\diff E}\Phi_\astro(k, \alpha) = k E^{-\alpha}.
\end{equation}
Here, $k$ is a normalization constant and $\alpha \geqslant 2$ is the spectral index of the spectrum.
We fix the spectral index at $\alpha = 2.0$ as suggested by Fermi shock acceleration in the jets and cores of high energy astrophysical object, and as determined from preliminary fits to the IC data \cite{Aartsen:2013jdh}.\footnote{Although, as mentioned in \citep{Aartsen:2013jdh}, if the lack of events at higher ($\geqslant 1.2$ PeV) energies are taken into account, the best fit for the spectral index indicates a softer spectrum, \ie, $\alpha=2.3$ (see also \cite{Anchordoqui:2013qsi}). As the event statistics are presently too low to make a definite conclusion either way, we will assume $\alpha = 2.0$ for the rest of the paper.}
To evaluate the total event rate expected at IC, we sum the total neutrino flux for each flavor at Earth from \dm\ decay and astrophysical sources, \ie,
\ifonecol
\begin{equation}
\frac{\diff}{\diff E}\Phi^{\lambda}_\text{Total}\left(m_\dm, \tau_\dm, k\right) = \frac{\diff}{\diff E}\Phi^{\lambda}_{\dm}\left(m_\dm, \tau_\dm \right) + \frac{\diff}{\diff E}\Phi^{\lambda}_\text{Astro}\left (k, \alpha \right )\Big\vert_{\alpha = 2},
\end{equation}
\else
\begin{multline}
\frac{\diff}{\diff E}\Phi^{\lambda}_\text{Total}\left(m_\dm, \tau_\dm, k\right)\\
= \frac{\diff}{\diff E}\Phi^{\lambda}_{\dm}\left(m_\dm, \tau_\dm \right) + \frac{\diff}{\diff E}\Phi^{\lambda}_\text{Astro}\left (k, \alpha \right )\Big\vert_{\alpha = 2},
\end{multline}
\fi
where, $\Phi^{\lambda}_\text{Total}$ indicates the total flux at Earth for the neutrino of flavor $\lambda$, with $\lambda = e\text{, }\mu\text{, or }\tau$.

The IceCube has recently carried out a survey of very high energy contained events, \ie, both cascades and muon track events with their starting vertex located within the body of the detector.
Based on the 28 events detected over a run-time of 662 days, the IC finds the best-fit to the astrophysical $E^{-2}$ flux to be
\begin{equation}\label{eqn:icfitflux}
E^{2}\frac{\diff \Phi_\ic}{\diff E} = 1.2 \times 10^{-8} \flxunit.
\end{equation}
For our analysis, we use the maximal atmospheric background including 90\% charm limits reported by the IceCube Collaboration \cite{Aartsen:2013jdh}.
\newinsrt{Considering the limited statistics of the IC data, and the fact that we assume the most conservative signal after subtracting the maximal atmospheric background, the effect of systematic and statistical errors in the background would most likely only expand the available \dm\ parameter space.}
The statistical degree of match of this best-fit flux with the \edited{signal} is given by the \chisq: the smaller this value, the better the match.
For the IC best-fit flux $\Phi_\ic$, $\chisq = 10.7$.
Note that, while IC can only measure the energy deposited in the detector, we have conservatively assumed these energies are equal to the incident neutrino energies.
This is only strictly true for \nue\ charged-current events; for all other events the deposited energy represents the minimum energy of the incident neutrino (see also discussion in \cite{Laha:2013lka}).

Using the fluxes obtained above, along with the effective exposure areas for contained events over a period of 662 days of IC's run-time given in \cite{Aartsen:2013jdh} for each neutrino flavor, we calculate the total events expected for a given set of the parameters $m_\dm, \tau_{\dm}$, and $k$, for each of the possible decay modes of the \dm\ particle.

\subsection{\label{subsec:ltbounds}Constraints on the \dm\ parameter space}
\begin{figure*}[htb]
\centering
	\includegraphics[width=0.99\textwidth]{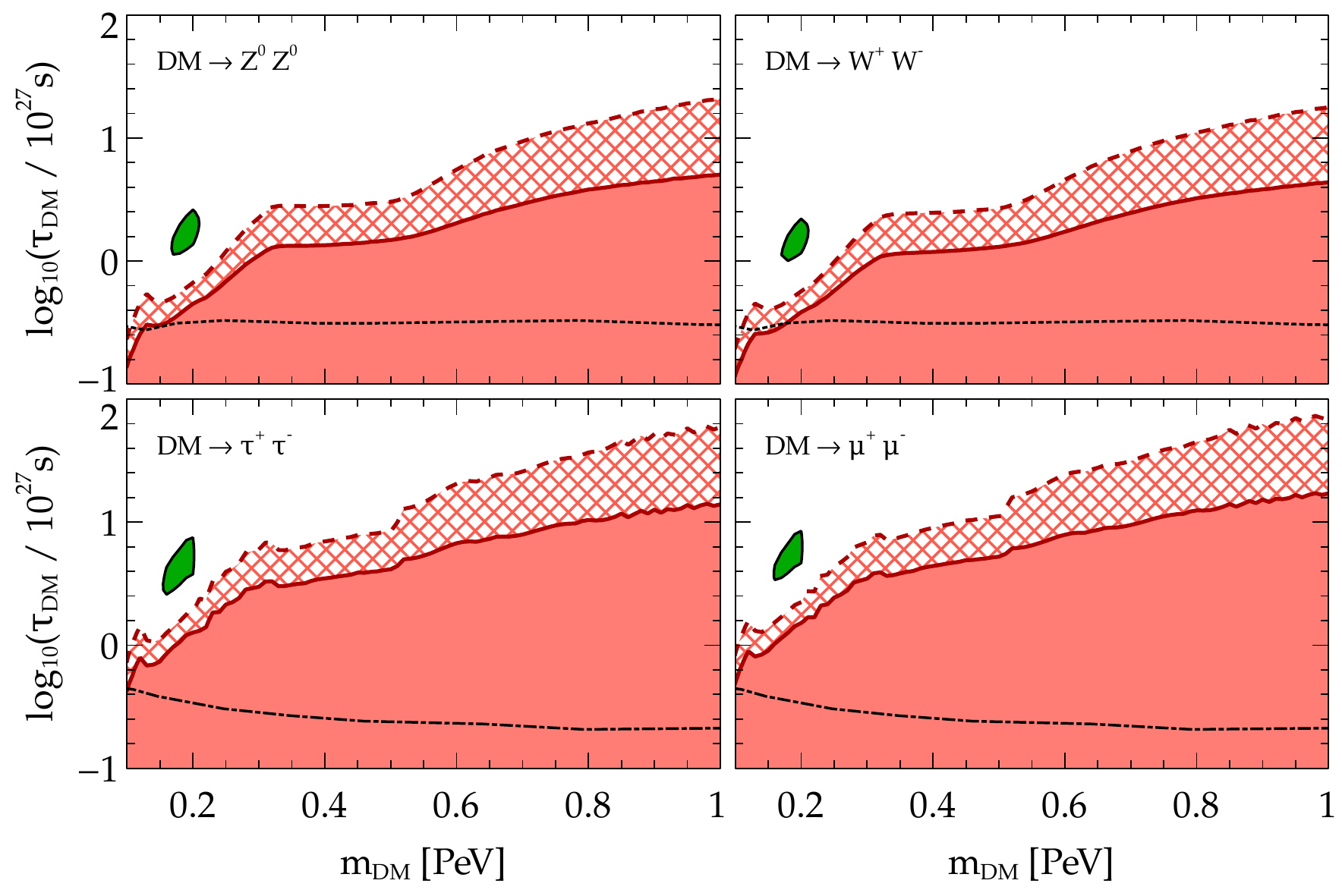}
	\caption{\label{fig:excl}Region of $m_\dm-\tau_\dm$ parameter space consistent with IC data if the total neutrino flux is a composition of flux from \dm\ decay and astrophysical flux at the IC best-fit (see eq.\ \eqref{eqn:icfitflux}).
	\edited{The hatched and red shaded regions are ruled out at 90\% and 99\% \cl\ respectively, while the green patch shows the region of parameter space consistent with data at $3\sigma$.}
	To compare with existing bounds on lifetimes from gamma ray observations, we show the bound obtained in \protect \cite{Murase:2012xs} when the \dm\ decays to a $W^{\pm}$ pair (black dotted curve) in the top panel plots ($\dm \to Z^{0}Z^{0}$ and $\dm \to W^{+}W^{-}$), and that obtained when the \dm\ decays to $\mu^{+}\mu^{-}$ (black dot-dashed curve) in plots in the bottom panel ($\dm \to \tau^{+}\tau^{-}$ and $\dm \to \mu^{+}\mu^{-}$), with the region below the curves excluded in both cases.}
\end{figure*}

To analyze the region of the $m_{\dm}-\tau_\dm$ parameter space, in view of the IC events, if the total flux reaching the detector is a combination of fluxes from \dm\ decay and astrophysical sources, we fix the astrophysical flux at the IC best-fit (and given by eq.\ \eqref{eqn:icfitflux}), and carry out a scan of the \dm\ parameter space by varying the mass and lifetimes in the range $100\text{ TeV} \leqslant m_\dm \leqslant 1\text{ PeV}$ and $10^{26}\text{ s} \leqslant \tau_\dm \leqslant 10^{29}\text{ s}$ respectively.
\edited{For each set of parameters we calculate the total event rates expected at the IC, and calculate the corresponding \chisq.
By doing this in turn for each \dm\ decay channel, we plot the parameter space and show the region of this space that is disfavored at 90\% and 99\% confidence levels (\cl) respectively.
As previously noted, the IC fit is in significant tension with the lower energy events (30 -- 100 TeV).
Physically, our analysis represents the scenario where the decay of an $\order{100\text{ TeV}}$ \dm\ particle augments the low energy events from the astrophysical flux to provide a significantly better match to the observed events, while the high energy events closer to the PeV are explained by the astrophysical flux alone.
It is, therefore, obvious that fixing the astrophysical flux at the IC best-fit tends to push the accessible \dm\ parameter space closer to the low $m_\dm$ regions --- the observed events at the high energies are either already consistent with those expected from the astrophysical flux alone or, as in the case of the two high energy bins immediately below 1 PeV, somewhat less, and added events from \dm\ decay only serves to increase the mismatch in such a scenario.
This is borne out by the allowed \dm\ parameter space at $3\sigma$, which lies in the region around $m_\dm \sim 200$ TeV.
For heavier \dm\ with mass $m_\dm \geqslant 500$ TeV, decay lifetimes are constrained to be more than $10^{27}$s at 90\% \cl}
The results are shown in figure \ref{fig:excl}.

\begin{figure*}[htb]
\centering
	\includegraphics[width=0.99\textwidth]{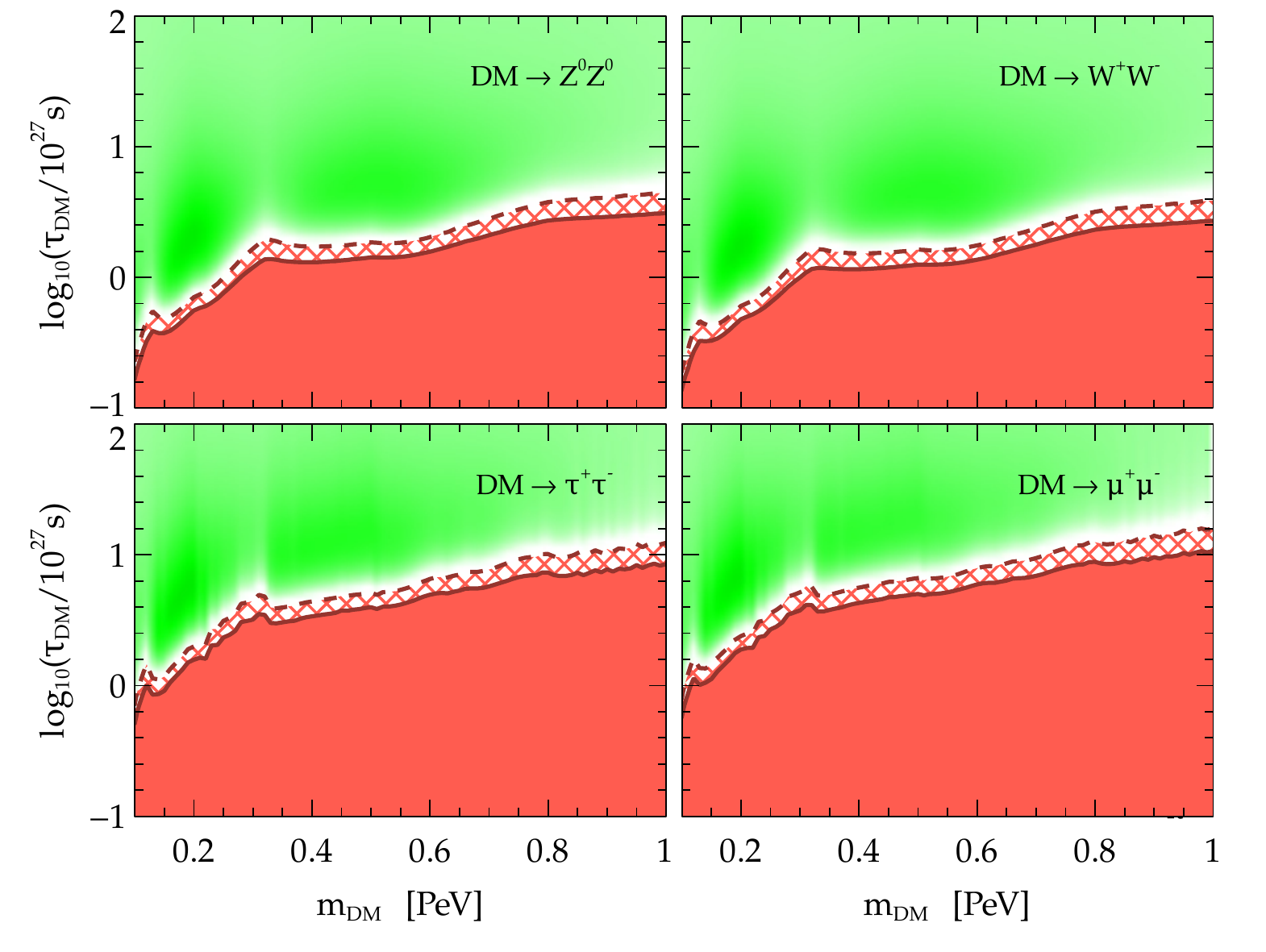}
	\caption{\label{fig:exclred}\newinsrt{Region of $m_\dm-\tau_\dm$ parameter space consistent with IC data if the total neutrino flux is a composition of flux from \dm\ decay and astrophysical flux with an undetermined normalization.
	For each set of values ($\tau_\dm$, $m_\dm$) in the parameter space, the astrophysical flux normalization is allowed to vary as a free parameter to obtain the minimized \chisq.
	The red hatched and shaded regions are ruled out at 90\% and 99\% \cl\ respectively, while the $3\sigma$ allowed region consistent with data in shown in green.}
}
\end{figure*}

\newinsrt{To illustrate the dependence on the astrophysical flux normalization, we also carry out a similar analysis by now letting the normalization to the astrophysical flux vary as a free parameter, in addition to the \dm\ mass and lifetime.
Specifically, we scan the $m_\dm-\tau_\dm$ parameter plane as above, only now for each coordinate in the plane, we calculate the minimum \chisq\ obtained by allowing the astrophysical flux normalization to vary as a free parameter in the range
\begin{equation}\label{eqn:astrolims}
	10^{-9} \flxunit \leqslant E^{2}\Phi \leqslant 10^{-7} \flxunit\,,
\end{equation}
as opposed to fixing it at the IC best-fit value.
By doing this for the entire $m_\dm - \tau_\dm$ parameter space, we plot the 90\% and 99\% \cl\ exclusion regions in figure \ref{fig:exclred}.
We see that allowing the astrophysical flux normalization to vary, the exclusion limits on the \dm\ parameter space are reduced significantly. This opens the parameter space to allow for heavier \dm\ masses with reasonable lifetimes, to within $3\sigma$ significance.
}

\subsection{\label{subsec:bfit}Best-fit parameters}
In this section, we derive the values for the \dm\ decay mass and lifetimes and the astrophysical flux normalization that best fits the IC data.

\begin{figure*}
\centering
	\includegraphics[width=0.99\textwidth]{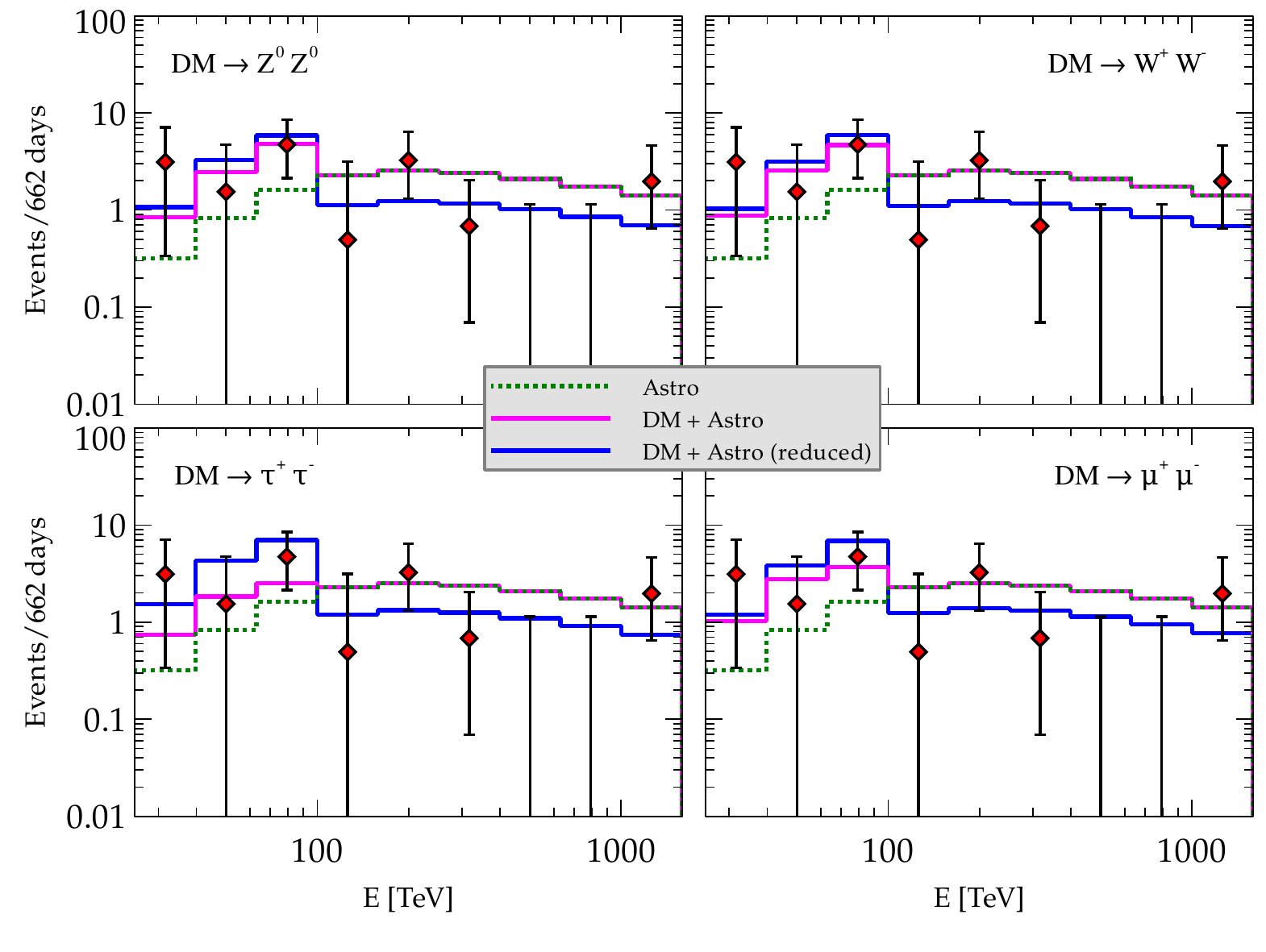}
	\caption{\label{fig:bfevents}Signal-only event rates (atmospheric backgrounds have been subtracted out) corresponding to the best fit parameters for each decay channel as listed in tables \ref{tab:bfpars} and \ref{tab:bfparsred} with the astrophysical flux normalization set at the IC best-fit (magenta lines) and at the best-fit obtained by doing a full three-parameter scan (blue lines) respectively.
	Event rates from the IC best-fit flux are also shown (green dotted lines), as are the actual event numbers seen at IC along with their associated errors (red dots).}
\end{figure*}

%To do so for each of the possible \dm\ decay channels, we carry out two different kinds of analyses:
%\begin{enumerate}
%\item
For each of the possible \dm\ decay channels we keep the astrophysical flux normalization fixed at the IC best-fit value, \ie, $k = E^{2}\diff\Phi_\text{IC}/\diff E = 1.2 \times 10^{-8} \flxunit$ and vary the \dm\ mass and lifetime over the following ranges
%\begin{subequations}\label{eqn:dmrange}
\begin{equation}\label{eqn:dmrange}
	100 \leqslant \left(\frac{m_\dm}{1\text{ TeV}}\right) \leqslant 1000\,,\qquad
%\end{equation}
%\begin{equation}
	1 \leqslant \left(\frac{\tau_\dm}{10^{26}\text{ s}}\right) \leqslant 1000
\end{equation}
%\end{subequations}
respectively.\footnote{Although $\tau_\dm$ does not have an observed or theoretically motivated upper bound, the neutrino flux from \dm\ decay falls with increasing decay lifetimes, and when as large as $10^{29}$s, it already leads to unobservably small event rates at IC.
Here, we set the upper bound for the $\tau_\dm$ parameter space scan to $10^{29}$s for computational purposes --- for the purposes of the analysis, taking even larger values of $\tau_\dm$ is equivalent to assuming the neutrino events seen at IC are solely due to the astrophysical power-law flux.}
Thus considering each decay channel in turn, we calculate the number of events expected due to a sum total of the astrophysical flux and that from the decay.
The resulting best fits and \chisq\ representing the degree of match are shown in table \ref{tab:bfpars}.
Event rates corresponding to the best-fit parameters are shown in figure \ref{fig:bfevents}.
It is evident from the figure that, especially, at the lower energies, \ie, $30\text{ TeV} \leqslant E \leqslant 100\text{ TeV}$, the combined \dm\ and astrophysical flux gives a better fit to the observed data than the IC best-fit astrophysical $E^{-2}$ flux.
\newinsrt{To provide a reasonable quantitative measure of the comparative goodness-of-fits for the \dm\ + Astro models vis-a-vis the IC best-fit for each of the four decay channels, we determine how significantly the former improves the match with data over the former by means of the F-test (see, \eg, \cite{Bevington}).
The p-values ($0 < p < 1$) obtained in F-tests are representative of the degree of rejection of the null hypothesis, which here refers to the proposition that the \dm\ + Astro model provides no improvement in the fit to the observed data in comparison to the IC best-fit.
The smaller the p-value, the stronger the degree of rejection of the null hypothesis is in favor of the chosen model.
As a rule of thumb, a p-value $p \leqslant 0.05$ indicates that the null hypothesis is \emph{strongly} disfavored \cite{Beringer:1900zz}.
For our case, the relative F-statistic between the \dm\ + Astro and IC best-fit models are tabulated in table \ref{tab:bfpars} for each of the decay channels.
Given the limited statistics of the present IC data, these results are indicative of the small, yet notable, improvement that the \dm\ + Astro fits show over the IC best-fit with respect to the observed events.}

We also similarly analyse the scenario where the astrophysical flux normalization is allowed to vary as a free parameter, in addition to the $m_\dm$ and $\tau_\dm$.
The results (also shown in figure \ref{fig:bfevents}) illustrate the improvement in match that a reduced astrophysical flux --- as shown in table \ref{tab:bfparsred}, the best fit normalization turns out to be approximately half the IC best-fit in each case --- gives with the high energy events, while still being consistent with events at the lower energies.
It is clear that, if neutrino fluxes arrive both from \dm\ decay and astrophysical sources, a reduced normalization for the latter is more consistent with the observation, including the high energy bins immediately below 1 PeV where no events are seen.
Significantly, an astrophysical flux with its normalization at the IC best-fit seems to be in some disagreement with these ``gaps'' in the event spectrum.
\newinsrt{The improvement of the fit with observed data in this case is also notable by the drop in p-values from the F-test, with respect to those obtained in table \ref{tab:bfpars} for the case of the \dm\ + Astro model with the IC astrophysical flux.
}

The value of the spectral index in these analyses is determined by the event spectrum at the higher energies ($\geqslant 200$ TeV), where the astrophysical flux is the sole contributor to the incident neutrino flux.
As such the best match to the events in these energies comes from an incident astrophysical flux that goes as $E^{-2}$, and we have, therefore, chosen to fix the $\alpha$ at this value for our analyses.
Softer flux spectra with $\alpha \geqslant 2$ would lead to event rates that fall off quickly and, for large enough $\alpha$, become too small to explain the two events seen at energies above 1 PeV.
If the incident neutrino flux had a softer spectral index not very different from $\alpha = 2$ (so that it were still consistent with the high energy events), the lower energies, where fluxes from both \dm\ decay and the astrophysical power-law spectrum contribute, would witness a slight increase in the relative contribution of the latter over the former, and, a similar analysis would, consequently, yield correspondingly larger best-fit \dm\ lifetimes.

\begin{table}[htb]
\centering
%\begin{ruledtabular}
\begin{tabular}{|l c c c c|}
\hline
Decay mode & 
$m^\text{b.f.}_\dm\,(\text{TeV})$ &
$\tau^\text{b.f.}_{\dm}/10^{27}\text{s}$ & 
$\chisq$ &
p-value
\\
\hline
%\noalign{\smallskip}
$\dm \to Z^{0}Z^{0}$
& 196.69 
& 1.77
& 8.861
& 0.268
\\
$\dm \to W^{+}W^{-}$
& 195.83
& 1.50 
& 8.864
& 0.268
\\
$\dm \to \tau^{+} \tau^{-}$
& 148.49
& 7.21 
& 9.554
& 0.390
\\
$\dm \to \mu^{+} \mu^{-}$
& 166.01
& 5.20 
& 8.976
& 0.282
\\
\hline
\end{tabular}
\caption{\label{tab:bfpars}The best fit parameters for each of the \dm\ decay modes and a comparison with the IC best-fit power law spectrum.
\newinsrt{The rightmost column represents the p-values for the F-test to determine if, in comparison to the IC best-fit, the improvement to the fit obtained in the \dm\ + Astro model is significant.
A lower p-value indicates a more significant improvement.}
The astrophysical flux normalization is fixed at the IC best-fit (see eq.\ \eqref{eqn:icfitflux}).}
%\end{ruledtabular}
\end{table}

\begin{table}[htb]
\centering
%\begin{ruledtabular}
\begin{tabular}{|l c c c c c|}
\hline
Decay mode & 
$m^\text{b.f.}_\dm\,(\text{TeV})$ & 
$\tau^\text{b.f.}_{\dm}/10^{27}\text{s}$ &
$\mbox{\small $\frac{E^{2}\Phi^{b.f.}_\text{Astro}}{(10^{-9} \text{\flxunit})}$}$ &
$\chisq$ &
~~p-value~~
\\
\hline
%\noalign{\smallskip}
$\dm \to Z^{0}Z^{0}$
& 191.82
& 1.27
& 5.86
& 4.209
& 0.061
\\
$\dm \to W^{+}W^{-}$
& 199.32
& 1.18
& 5.82 
& 4.209
& 0.061
\\
$\dm \to \tau^{+} \tau^{-}$
& 176.61
& 3.11
& 6.29
& 4.188
& 0.060
\\
$\dm \to \mu^{+} \mu^{-}$
& 197.97
& 5.01 
& 6.14
& 4.445
& 0.072
\\
\hline
\end{tabular}
%\end{ruledtabular}
\caption{\label{tab:bfparsred}The best fit parameters for each of the \dm\ decay modes and a comparison with the IC best-fit power law spectrum.
In addition to the \dm\ mass and lifetime, the normalization for the astrophysical flux is varied as a free parameter.
\newinsrt{As in table \ref{tab:bfpars}, the p-values indicate the significance of improvement to the fit with the \dm\ + Astro model over the IC best-fit.
The low values of the p-value indicate that the fit to the data in the \dm\ + Astro model with  a reduced astrophysical flux improves upon the IC best-fit significantly.
Conventionally, $p \leqslant 0.05$ indicates strong presumption against the null hypothesis, which in this case refers to the hypothesis that the fit \emph{does not} improve statistically significantly.
Furthermore, a comparison of the p-values obtained here with those in table \ref{tab:bfpars} reveals that a the \dm\ + Astro model fits the IC data better with a reduced astrophysical flux normalization.
}
}
\end{table}

%\begin{figure*}
%\centering
%	\begin{subfigure}[b]{0.48\textwidth}
%		\includegraphics[width=\textwidth]{figures/ZZEvents}
%	\end{subfigure}
%	~
%	\begin{subfigure}[b]{0.48\textwidth}
%		\includegraphics[width=\textwidth]{figures/WWEvents}
%	\end{subfigure}
%	\\
%	\begin{subfigure}[b]{0.48\textwidth}
%		\includegraphics[width=\textwidth]{figures/TauTauEvents}
%	\end{subfigure}
%	~
%	\begin{subfigure}[b]{0.48\textwidth}
%		\includegraphics[width=\textwidth]{figures/MuMuEvents}
%	\end{subfigure}
%	\caption{\label{fig:bfevents}Signal-only event rates (atmospheric backgrounds have been subtracted out) corresponding to the best fit parameters for each decay channel as listed in tables \ref{tab:bfpars} and \ref{tab:bfparsred} with the astrophysical flux normalisation set at the IC best-fit (magenta lines) and at half its value (blue lines) respectively.
%	Event rates from the IC best-fit flux are also shown (green lines), as are the actual event numbers seen at IC along with their associated errors (red dots).}
%\end{figure*}

\section{\label{sec:conc}Conclusion}
We have shown that if the events seen at the ultra-high energies at the IceCube are understood to be due to a combination of neutrino fluxes from both \dm\ decay and astrophysical sources, the observations permit a \dm\ mass in the range 150--250 TeV, with decay lifetimes of \order{10^{27}\text{s}}.
By considering important decay pathways of the \dm\ particle, \viz,
\begin{inparaenum}[\itshape i\upshape)]
	\item $\dm \to Z^{0} Z^{0}$,
	\item $\dm \to W^{+} W^{-}$,
	\item $\dm \to \tau^{+} \tau^{-}$, and
	\item $\dm \to \mu^{+} \mu^{-}$,
\end{inparaenum} we have done a largely model independent scan of the $m_\dm-\tau_\dm$ parameter space to check for consistency with IC data.
We see that, while, as expected, the channel $\dm \to \mu^{+}\mu^{-}$ is the most strongly constrained of the four, the recent events still allow for consistency with a significant region in the parameter space around $m_\dm \sim 200$ TeV and reasonable lifetimes for all four channels.

Despite the limited statistics that the observation of only a score of high energy events entails, we see that a total neutrino flux with contributions from \dm\ decay and astrophysical power law spectrum provides for a better match to the observation than that from an astrophysical source only flux, no matter what spectral shape one assumes for the latter.
Especially at the low energies around 30 -- 100 TeV, where the best-fit astrophysical flux from the IC analysis predicts event rates consistently less than that observed, the contribution of the neutrino flux from the decay of a $\sim 200$ TeV \dm\ species is significant and serves to significantly enhance the consistency between theory and experiment.

On the other hand, a neutrino flux augmented by \dm\ decay allows a significantly reduced astrophysical flux to still be consistent with the low energy events, while, concurrently inducing notable improvement in consistency with events at the high energy ($E_{\text{mean}} \geqslant 400$ TeV) bins.
By allowing the astrophysical flux normalization vary as a free parameter, in addition to the \dm\ mass and lifetime, we show, that a reduced $E^{-2}$ astrophysical flux at roughly half the IC best-fit provides a much better match with observations, including being consistent with the lack of events in some of the high energy ($E_\text{mean} \sim 500\text{ and }800$ TeV) bins, to within statistical errors.
\edited{The improvement in consistency with observations in this case is also clear from the drop in the p-value of the relative F-test, showing that the improvement in the goodness-of-fit over the IC best-fit is statistically significant.}

As the recent events at the IC show, the existence of an astrophysical flux of neutrinos looks increasingly true.
However, notwithstanding the low statistics thus far, the IC events do seem to indicate that a simple power law spectrum for the astrophysical flux cannot, by itself, explain these observations completely.
Accumulation of events in the IC as it collects data in the near future will serve as an important test for the consistency of the observations with the astrophysical flux predicted from standard calculations.
If, akin to that seen with the first set of UHE events, inconsistencies between those predictions and observations continue to persist, it will perhaps bring into consideration the existence of additional component(s) to the neutrino flux arriving at the Earth.
At these remarkably high energies, one of the (rather small set of) possible neutrino sources is from the interactions, either decay or annihilation, of a dark matter species with its mass \order{100\text{ TeV}}.
While \dm\ annihilation would not contribute a statistically observable number of events at the IC, decays to standard model particles with lifetimes \order{10^{27}\text{ s}} lead to detectable neutrino fluxes.
Given the limited statistics at present, we have shown that a combination of neutrino fluxes from \dm\ decay and astrophysical sources proves to be more consistent with observations than an astrophysical power-law spectrum alone.
We have also shown, in addition, that it might well be possible that the magnitude of the actual astrophysical flux seen by the IC detector might be as low as one-half the present best-fit value.
While, if the only component in the neutrino flux were astrophysical in nature, such a low flux would have been too small to explain the low energy events, the contribution of an additional neutrino flux from \dm\ decay, which becomes pertinent precisely at these low energies, would be enough to make up for the deficit.
In the process, we have shown that a UHE neutrino flux composed of neutrinos from \dm\ decay, in addition to those from an astrophysical power-law spectrum with a significantly reduced magnitude, is in good agreement with present observations.

With more data and consequently improved statistics in the future, the IC should be in prime position to conclusively determine if it is indeed seeing events from \dm\ decays in addition to those from astrophysical sources or, if, like many analyses based on low statistics made prior to this, the present conclusions may get modified.
To determine either way, and to distinguish the signal from a flux component made up of neutrinos from \dm\ decays against a pure astrophysical power-law flux, however, it would be important to note the subtle changes to the shape of the event spectrum that would correspondingly occur.

\begin{acknowledgments}
We are grateful to Ty DeYoung for providing us with detailed description of the statistical analysis method, the F-test, used in this paper.
In addition, AB would like to thank Raj Gandhi and Ranjan Laha for useful discussions.
This research was supported by US Department of Energy contracts DE-FG02-91ER40664, DE-FG02-04ER41319, DE-FG02-04ER41298, DE-FG03-91ER40662, DE-FG02-13ER41976 and DE-SC0010114.
\end{acknowledgments}

\bibliographystyle{JHEP}
\bibliography{DM_at_IC0}

\end{document}